\documentclass[conference,10pt]{IEEEtran}

\IEEEoverridecommandlockouts
\usepackage{balance}
\usepackage{mathptmx} 
\usepackage{cite}
\usepackage{amsmath,amssymb,amsfonts}
\usepackage{algorithmic}
\usepackage{graphicx}
\usepackage{textcomp}
\usepackage{xcolor}
\usepackage{array}
\usepackage{makecell}
\usepackage{lipsum}
\usepackage{enumitem}
\usepackage{ifthen} 
\usepackage{comment}
\usepackage{fancyhdr}
\usepackage{tabularx}
\usepackage[    protrusion=true,
            expansion=true,
            final,
            babel
                ]{microtype}

\usepackage{multicol}
\usepackage{caption}
\usepackage{kantlipsum}

\newcommand{\ALTIS}{Altis}

\newcommand{\ttt}{\texttt}

\newcommand{\paragraphbe}[1]{\vspace{.75ex}\noindent{\bf \em #1} }

\newboolean{publicversion}
\setboolean{publicversion}{false}

\ifthenelse{\boolean{publicversion}}{
  \newcommand{\grumbler}[2]{}
  \widowpenalty=1000
}{
  \newcommand{\grumbler}[2]{\textcolor{red}{\bf #1: #2}}
}

\ifthenelse{\boolean{publicversion}}{
  \newcommand{\mybad}[2]{}
  \widowpenalty=1000
}{
  \newcommand{\mybad}[2]{\textcolor{blue}{\bf #1: #2}}
}

\def\BibTeX{{\rm B\kern-.05em{\sc i\kern-.025em b}\kern-.08em
    T\kern-.1667em\lower.7ex\hbox{E}\kern-.125emX}}
    
\date{\vspace{-5ex}}
\begin{document}

\title{\ALTIS{}: Modernizing GPGPU Benchmarks\\
}

\author{\IEEEauthorblockN{Bodun Hu}
\IEEEauthorblockA{\textit{University of Texas at Austin} \\
Austin, USA \\
bodunhu@utexas.edu}
\and
\IEEEauthorblockN{Christopher J. Rossbach}
\IEEEauthorblockA{\textit{University of Texas at Austin and VMware Research Group} \\
Austin, USA \\
rossbach@cs.utexas.edu}

}
\pagestyle{plain}
\maketitle

\begin{abstract}
This paper presents \ALTIS, a benchmark suite for modern GPGPU computing. Previous benchmark suites such as Rodinia\cite{5306797} and SHOC\cite{Danalis:2010:SHC:1735688.1735702} have served the research community well, but were developed years ago when hardware was more limited,  software supported fewer features, and production hardware-accelerated workloads were scarce. Since that time, GPU compute density and memory capacity has grown exponentially, programmability features such as unified memory, demand paging, and HyperQ have matured, and new workloads such as deep neural networks (DNNs), graph analytics, and crypto-currencies have emerged in production environments, stressing the hardware and software in ways that previous benchmarks did not anticipate. Drawing inspiration from Rodinia and SHOC, \ALTIS{} is a benchmark suite designed for modern GPU architectures and modern GPU runtimes, representing a diverse set of application domains. By adopting and extending applications from Rodinia and SHOC, adding new applications, and focusing on CUDA platforms, \ALTIS{} better represents modern GPGPU workloads to enable support GPGPU research in both architecture and system software.
\end{abstract}

\section{Introduction}
GPUs have become popular hardware accelerators in many computing domains. Originally used primarily for 3D rendering, their use as General Purpose Graphics Processing Units (GPGPUs) has grown rapidly.
Heterogeneous computing using both CPUs and GPUs is now the dominant practice in many domains such as ML~\cite{nextplatform-GPU-medical,gpu-medical}, graph~\cite{thinci,thinci-eetimes,nvgraph-doc},
and video analytics~\cite{nvidia-selfdriving}, and crypto-currencies~\cite{blockchain18}. At the same time, dramatic improvements in programmability for GPGPU workloads have been driven by rapidly evolving high-level languages and runtimes such as CUDA~\cite{CUDA}.
Unfortunately, current GPGPU benchmark suites have not co-evolved with the hardware, programming frameworks and runtimes, or production workloads, so their ability to drive relevant research findings has become limited. 

Existing benchmark suites such as Rodinia~\cite{5306797} and SHOC~\cite{Danalis:2010:SHC:1735688.1735702} were designed to enable researchers to better understand the characteristics and demands of heterogeneous systems. Applications were curated such that each benchmark exhibits unique behaviors to exercise important GPU subsystems and components. This allows architects to design better hardware and enables system software developers to improve efficiency and programmability. However, current GPGPU benchmark suites have not kept up with the evolution of hardware and programming frameworks: in general, they do not use or stress newer features introduced over multiple generations of CUDA, such as HyperQ, dynamic parallelism, NVLink~\cite{nvlink}, preemptive context switch~\cite{isca-2014-preemptive}, and UVM\cite{CUDA}. While the hardware has evolved to provide more raw computing power, memory capacity, and memory bandwidth, the task of scaling benchmarks to adapt to that growth has either been left to the user, (e.g. with Rodinia), or entirely neglected (SHOC). 

\begin{figure}
    \centering
    \includegraphics[width=\columnwidth]{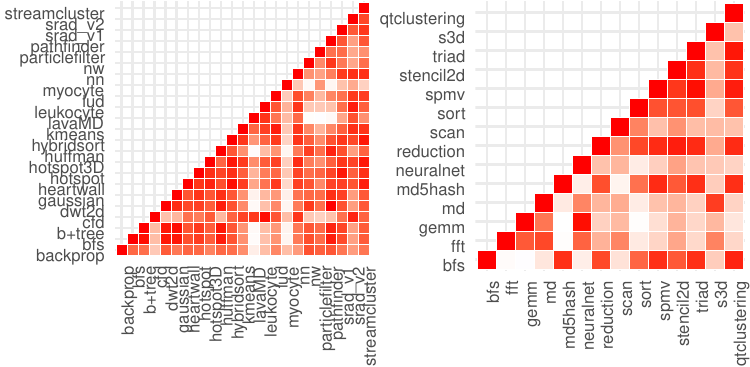}
    \caption{The Pearson correlation matrix for Rodinia (left) and SHOC (right). The \ttt{x} axis represents var1 and the \ttt{y} axis represents var2. Darker colors represent greater positive correlation value.}
    \label{fig:corr1}
\end{figure}

GPUs are the \emph{de facto} platform of choice for training deep neural networks. Frameworks like Tensorflow\cite{tensorflow2015-whitepaper} and PyTorch\cite{paszke2017automatic} are widely adopted in the research and industry environments. While these platforms are available to researchers, they are not designed to study hardware behaviors or system software design. Targeted toward production environments, they introduce large memory footprints, making them difficult for hardware architects using simulators. DNNs have become arguably the most important domain driving GPGPU technology trends, but the minimal neural network workloads in SHOC and Rodinia are not representative of the current state of the art in DNN models or algorithms.

We present \ALTIS{} to address emerging needs of GPGPU researchers working in both architecture and systems. \ALTIS{} preserves what is useful in current suites, extending them to use new features and exploit new hardware capabilities. \ALTIS{} introduces new workloads and feature support to reflect the evolution of the GPGPU application landscape. We limit our focus to CUDA, unlike previous suites which also support OpenCL or other frameworks, because CUDA has become the dominant platform in production settings, and because production OpenCL support for advanced features like demand-paging has been slow to emerge. We make the following contributions:
\begin{itemize}
    \item We demonstrate areas in which existing benchmark suites no longer meet researchers' needs, specifically in workload diversity, problem sizes, and programming features.
    \item We demonstrate improvements made by \ALTIS{} and show capabilities of emerging GPUs more comprehensively. We implement support for new features up to CUDA 10.0.
    \item We introduce new benchmarks to characterize emerging applications domains,  emphasizing  neural networks by adding commonly used DNN kernels to \ALTIS{}. 
\end{itemize}

\section{Background and Related Work}
\label{sec:related}

\paragraphbe{Rodinia}~\cite{5306797} is a suite of applications designed for heterogeneous systems released in 2009. It comprises applications representing different behaviors of the Berkeley dwarfs~\cite{dwarfs}. The dwarfs are 13 major categories of computation conjectured to describe most types of parallel problems. Rodinia relies on that conjecture to comprise a diverse set of applications that include both communication and synchronization. However, Rodinia does not take advantage of new features like unified memory, HyperQ, Cooperative Groups, and Dynamic Parallelism~\cite{newCudaFeature}, because it still depends on CUDA 4.0. It does not utilize programming constructs or architectural support introduced in newer CUDA versions that can significantly improve performance: for example, new hardware capabilities such as half precision and tensor core computation are unused.

\begin{figure}
    \centering
    \includegraphics[width=0.8\columnwidth]{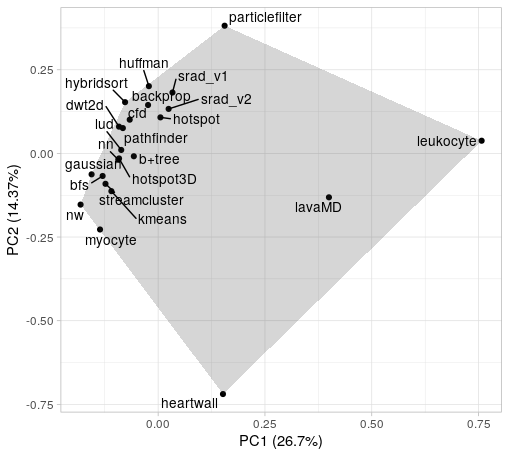}
    \caption{Rodinia PCA}
    \label{fig:rodinia_pca}
    \vspace{-.4cm}
\end{figure}

To analyze how well Rodinia exercises and stresses GPUs, we profiled performance for each benchmark using \ttt{nvprof}\cite{nvprof}. The tool reports metrics that quantify utilization of a component relative to  peak on a scale of 0 to 10: 0 is idle, 10 is full utilization. Many applications run multiple kernels: for these cases, we collect average utilization rate for all kernels, and report the maximum of those averages. To characterize how diverse applications in the suite are, we apply Principle Component Analysis (PCA) to the top 69 metrics supported by \ttt{nvprof} that contribute most to dissimilarity across all benchmarks in the suite using the default \ttt{run} configuration.
The metrics included are detailed in \autoref{tab:metrics} and methodology is described in \S\ref{sec:eval}. Our goal is to assess similarity/dissimilarity among benchmarks and show how well they exercise different components of the GPU and its runtime. The result is shown in~\autoref{fig:rodinia_pca}. The first three PCs represent 55\% of total variance, which accounts for more than half of the diversity across all benchmarks. A correlation matrix is shown in \autoref{fig:corr1}: an ideal matrix would be dark along the diagonal but light elsewhere, and would demonstrate that applications have different behaviors and exercise the system differently. However, 41\% and 70\% of applications have correlation values greater than $0.8$ and $0.6$ respectively, showing most applications in Rodinia are highly correlated with a few outliers. An improved suite that in which applications are more diverse is urgently needed.

\begin{figure*}
 \centering

 \begin{minipage}{.05\textwidth}
 \vspace{-.5cm}
  \includegraphics[width=\textwidth]{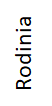} 
  \end{minipage}
 \begin{minipage}{.92\textwidth}
 \centering
 \includegraphics[width=\textwidth]{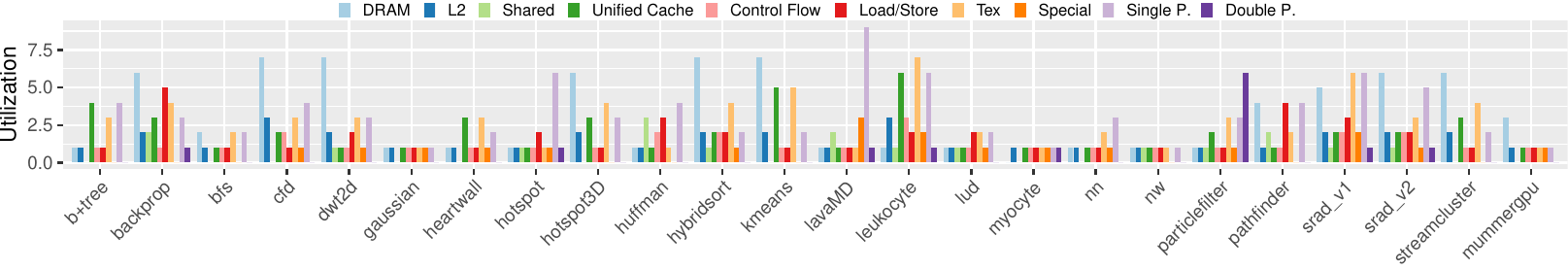}
 \end{minipage}
 \begin{minipage}{.05\textwidth}
 \vspace{-.5cm}
  \includegraphics[width=\textwidth]{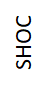} 
  \end{minipage}
 \begin{minipage}{.92\textwidth}
 \centering
  \includegraphics[width=\textwidth]{SHOC.pdf}
  \end{minipage}
  \caption{GPU resource utilization for Rodinia and SHOC.}
  \label{fig:rodinia}
  \label{fig:shoc}
\end{figure*}

\autoref{fig:rodinia} shows the average utilization rate and their standard deviation of the different
functional components and the memory hierarchy for each
application in Rodinia. 
A key observation is that many components have low utilization; even those comprising multiple kernels. Moreover, average GPU component utilization looks \emph{very similar} for many benchmarks, such as \texttt{gaussian}, \texttt{huffman}, \texttt{nw}, and \texttt{myocyte}. Many hardware components are not stressed to achieve maximum utilization, stifling research into improving those components.

\begin{figure}[]
    \centering
    \includegraphics[width=0.8\columnwidth]{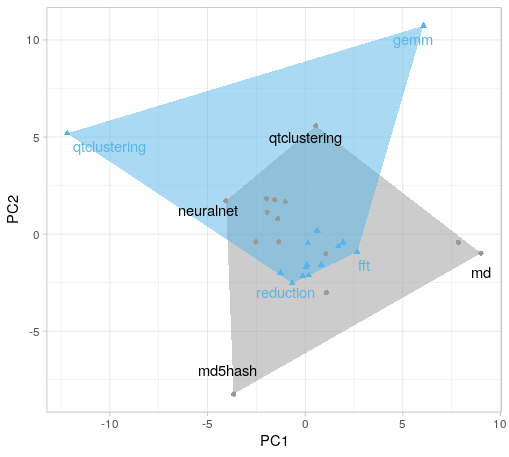}
    \caption{SHOC PCA. The black dots represent results from running the smallest data size and the red ones represent the benchmarks using the largest data size.}
    \label{fig:shoc_pca}
\end{figure}

\paragraphbe{SHOC.} 
Developed in 2010, SHOC\cite{Danalis:2010:SHC:1735688.1735702} differs from Rodinia by dividing workloads into two primary categories: stress/functional tests and performance tests. Stress tests use computationally demanding kernels to identify devices with bad memory, insufficient cooling, and other device component problems. Performance tests, on the other hand, focus on synthetic kernels and a handful of common parallel algorithms.
SHOC applications run in a framework which accepts user-specified parameters (e.g. number of iterations), and metrics, such as floating point operations per second, are recorded. 

Many programs in SHOC are basic parallel algorithms, reflecting an ever-shrinking subset of algorithms actually used in more modern applications. SHOC covers a variety of the Berkeley dwarfs, but does not capture the evolution or complexity of modern applications. Like Rodinia, SHOC was written when software and hardware features were more limited.

\autoref{fig:corr1} includes a correlation matrix for SHOC that shows overall less correlation than Rodinia (12\% and 31\% values over $0.8$ and $0.6$ respectively), but a handful of benchmarks are \emph{very} highly correlated with most others (e.g. \texttt{scan} and \texttt{neuralnet}). However, while SHOC suffers less from lack of diversity than Rodinia, it significantly under-utilized modern hardware. 
\autoref{fig:shoc} shows GPU resource utilization and standard variance for each application in SHOC, measured using the largest preset data size available. Unlike Rodinia, the utilization rate no longer exhibits a fixed pattern but varies over a wide range. This is because SHOC consists of microbenchmarks that target specific hardware components. Importantly, most components are not fully exercised to maximum capacity.  The PCA result is shown in \autoref{fig:shoc_pca} based on the same metrics used above, profiled with both the smallest and the largest predefined data sizes.  Like the PCA analysis for Rodinia (\autoref{fig:rodinia_pca}), with the exception of a very small number of outliers, workloads are mostly tightly clustered together. Tight clustering suggests that the workloads do not exercise the GPU in significantly different ways. As the data size increases, the workloads become even more clustered, showing that a key technology trend (increased memory capacity) is making these workloads even less useful for capturing a variety of important behaviors. This highlights the need for workloads to better cover the underlying metric space for modern and future GPGPU hardware and applications.

\paragraphbe{Other Benchmarks.}
More recent benchmarks such as Lonestar\cite{4919639}, Pannotia\cite{6704684}, and LonestarGPU~\cite{6983052} have addressed irregular parallelism, rather than stressing overall heterogeneous performance. Not all new CUDA features are supported. Sarita, Alsop, and Sinclair\cite{8167781} focus on benchmarking the effects of different levels of synchronization (from coarse to fine), but do not address trends in GPGPU software and hardware evolution. Parboil\cite{Stratton2012ParboilAR} provides a set of throughput computing applications useful for studying the performance of throughput computing  architecture  and  compilers. The MAFIA framework\cite{Jog:2015:AGM:2818950.2818979} is designed to target multi-application execution in GPUs. These efforts are motivated by similar challenges to GPGPU computing research: lack of workloads that actually exercise the features under test. Unfortunately MAFIA remains tied to older versions of CUDA by the need to support an in simulator implementation of the runtime, which both stifles co-evoluation, and limits the fidelity with which it can capture real system behaviors.  Chai~\cite{chai} address many of the same concerns motivating \ALTIS{} and incorporates new benchmarks up to CUDA 8.0, but focuses more on task and data partitioning rather than emerging  domains such as DNNs and relational algebra. Unlike \ALTIS{}, Chai's introduction of a new programming front end limits its ability to keep up with CUDA evolution.

Due to the rapid growth of popularity in machine learning, there has been significant focus on characterizing DNN behavior on GPUs~\cite{8573476}\cite{8695671}\cite{8573521}. Popular frameworks such as Tensorflow include primitive tools for users to analyze the computational demands of their models. Tango \cite{karki2019tango} is a framework to study behaviors of specific neural network model. MLPerf is benchmark suite for measuring the performance of machine learning (ML) software frameworks, ML hardware platforms  and ML cloud platforms. These systems could help GPGPU researchers, but because they focus on high-level and end-to-end behaviors, lack of control over scale and poor visibility into system-level and architecture-level phenomena make them hard for systems and architecture researchers to use for GPGPU research. A better benchmark suite targeted to this audience is long overdue.

\section{Motivation and Goals}

\subsection{Represent Emerging GPGPU Application Domains}
GPUs are used in many new domains not represented by workloads in Rodinia or SHOC, e.g., data analytics, graph processing, relational algebra, and DNNs.
While GPUs are the standard for DNN training, benchmark suites either do not include neural network based kernels, or include minimal workloads based on outdated techniques. Our goal is to include emerging domains, while preserving the ability of researchers to study low level behaviors of system software and individual kernels to explore improvements at those stack layers.

\subsection{Flexible Dataset Sizes}
A key aspect of existing benchmark suites that requires significant improvement is management of dataset and working set sizes. SHOC
provides four preset data sizes. This lack of flexibility makes it hard for SHOC to stay
relevant in the future, as advancing technology will eventually
cause even the largest data size to be too small to stress
GPU resources. Rodinia has the 
opposite problem, where benchmarks have no preset data size,
forcing the user to specify problem sizes, and making research results based on the suite difficult to interpret and compare. Users
must run data generation scripts even if they do not know what input size may be appropriate for the platform being measured. 
\ALTIS{} is designed for forward compatibility by supporting both updated 
default sizes and tooling to create arbitrary data sizes. A consequence of user-parameterizable data sizes is that real world datasets are not always possible. New data sets are synthesized. The user is given the freedom to specify the dataset size.
While \ALTIS{}'s support for modern default sizes and flexible non-defaults avoids 
shortcomings of SHOC (no flexibility) and Rodinia (no dataset size guidance),
the problem of updating defaults for future hardware is fundamental. In future work, we plan to explore providing feedback to help the user choose new default sizes based on utilization.

\subsection{Focus on CUDA and Recent CUDA Versions}
Supporting multiple GPGPU programming frameworks is a goal with considerable 
conceptual appeal, but one which increases maintenance burden and engineering effort
involved in keeping pace with hardware and software evolution. Because 
CUDA has become overwhelmingly the dominant framework for GPGPU, we limit
our focus to CUDA and on maintaining feature parity with recent releases.
In addition to performance improvements, each new release version of CUDA typically supports new programming features that can be used to more easily write more efficient code. It is essential to include these new features to understand their impact on performance, particularly as it is usually newer features that are of interest to systems researchers and architects. 

\section{The \ALTIS{} Benchmark Suite}

\emph{\ALTIS{} is available at: \href{https://github.com/utcs-scea/Altis}{https://github.com/utcs-scea/Altis}.}

In \ALTIS{}, like SHOC, benchmarks are divided into levels. Each level represents benchmarks whose behaviors of interest range from low level characteristics such as memory bandwidth to end-to-end performance on real world applications. While determining a set of benchmarks for \ALTIS{}, consideration was given to both the Berkeley dwarfs and emerging application domains such as analytics and DNNs. Our approach is to draw representative subsets from Rodinia and SHOC and augment those with workloads that represent emerging domains.

First, \ALTIS{} integrates a new set of benchmarks representing neural network layers commonly used in popular DNN models. Inclusion of DNN benchmarks creates a single suite that covers modern domains so that current research practice of cobbling together superset measurements from Rodinia, SHOC, Parboil, and others can be avoided. \ALTIS{} DNN workloads are parallelized with CUDA APIs and powered by libraries including cuBLAS and cuDNN (NVIDIA CUDA Deep Neural Network library). We use Darknet\cite{darknet13}, an open source Neural network framework, to construct neural networks. Darknet enables us to construct real-world models using existing building blocks. However, most of the kernels in Darknet do not utilize the cuDNN library, which causes poor performance compared to industrial standards like Tensorflow. Thus, we re-implement the most commonly used kernels with cuDNN, and remove extra memory operations to reduce memory footprints. Contrary to NN-focused suites like MLperf, which focuses on end-to-end measurements, \ALTIS{} isolates individual layers from DNN models, enabling finer-grained analysis. \emph{Critically, we believe that because \ALTIS{} uses cuDNN implementations, and extracts them from an end-to-end setting, it enables researchers to engage in detailed analysis based on state-of-the art code.}

\ALTIS{} aims to strike a balance between predetermined
    input sizes available in SHOC and customizable input sizes
    available in Rodinia. Benchmarks contains preset sizes
    optimized for systems with different compute capabilities,
    as well as a mechanism through which users can specify the
    size and other aspects of their input. This feature merges the favorable qualities from both Rodinia and SHOC. 

\paragraphbe{Characterizing new datasets.}
For all benchmarks adopted from Rodinia, users can either use the randomly generated data provided by \ALTIS{} tools, or the data originally provided in Rodinia (randomly generated). All data generated for additional benchmarks are randomly generated. While flexible sizing and random generation do mean \ALTIS{} benchmarks do not use ``real world'' datasets, we believe this is not a significant limitation for the type of architecture and systems research we envision it being used for. Importantly, while DNN research on end-to-end performance does require real world models and inputs to accurate reflect time-to-convergence for training, research on the behavior of individual kernels is not sensitive to the same concern.

While not all features are applicable in all workloads, \ALTIS{} includes support for each new CUDA feature \emph{in every workload where the feature is meaningful}. UVM and CUDA Event support are present in all workloads.
Dynamic parallelism and Cooperative Group support change the programming model sufficiently that new workloads (detailed below) are required to exercise them. HyperQ is meaningful only for workloads featuring kernels that actually have independent kernels that can be run concurrently without compromising application semantics. 
All benchmarks in \ALTIS{} support the most recent releases of CUDA (version 10 at the time of writing). Newly supported features are detailed below. 

\paragraphbe{Unified Memory} is a programmability feature supporting a shared address space across device across device and host, with demand paging support to eliminate the need for explicit data transfer code. When applications access data currently absent on the running device, the hardware automatically pages it in. This feature is supported in all \ALTIS{} workloads.

\paragraphbe{CUDA Events} enable accurate timing of functions and kernel calls. This is an improvement from previous suites which still use system time, risking low fidelity measurements, particularly in the presence of asynchrony. All \ALTIS{} workloads support CUDA events.

\paragraphbe{HyperQ} allows multiple independent CUDA kernels to execute in parallel on the same GPU if the resources are available. HyperQ uses 32 independent Work Distributor Queues to detect opportunities for parallelism, whereas old architectures uses a single Work Distributor Queue. We implement HyperQ support in all benchmarks that feature kernels that can execute independently (\texttt{DWT, LavaMD, SRAD}) and extend \texttt{Pathfinder} with a mode that runs independent but duplicate versions of the same kernels on different streams.

\paragraphbe{Dynamic Parallelism} enables currently executing CUDA kernels to call child CUDA kernels(nested parallelism). This feature is useful when running algorithms with hierarchical data structures and recursive algorithms with parallelism at each level. Previous suites do not feature workloads that use such idioms; we added the \texttt{Mandelbrot} workload to provide coverage for the feature. 

\paragraphbe{Cooperative Groups (Grid Sync)} provides another granularity of synchronization for kernel threads running on a GPU. GridSync allows users to sync all threads in the entire grid before beginning next section of computation. This features is useful for programs with disjoint phases of computation running right after one another. This provides finer synchronization granularity relative to previous CUDA versions that expose only \texttt{\_\_syncthreads()} to synchronize all threads in a single block. Like Dynamic Parallelism, using the feature effectively requires significant code change and is only advantageous for workloads that make frequent data-dependent kernel launches for which predicates would otherwise require data transfer. \ALTIS{} workloads that support it are \texttt{SRAD} and \texttt{kmeans}.
    
\paragraphbe{CUDA Graphs} present a new model for submitting jobs in CUDA. This features allow work to be defined as graphs instead of single operations. A graph consists of a series of operations, such as kernel launches and memory operations, defined as nodes. Connections between these nodes specify corresponding dependencies. This features enables launching multiple GPU operations with one CPU operation, hence reducing overheads. Graphs also enable the CUDA driver to perform optimizations because the whole workflow is visible.

\subsection{Level 0 Workloads}
Level 0 benchmarks are designed to measure low level
characteristics of the hardware. These benchmarks do the
simple task of measuring a single capability of the GPU and
therefore don’t represent any dwarfs or application domains.

\paragraphbe{BusSpeedDownload} measures the speed of the PCIe
bus by repeatedly transferring data of various sizes from the
host to the device. The data sizes are varied from 1kb to
500kb.

\paragraphbe{BusSpeedReadback} measures the speed of the
PCI bus, except in the opposite direction. Here, data is
transferred from the device to the host.

\paragraphbe{DeviceMemory} measures the bandwidth of different
components of the memory hierarchy on the device. This
includes global, constant, and shared memory.

\paragraphbe{MaxFlops} measures the maximum
achievable floating point operations per second on the
device. This is adopted from SHOC for single and double precision arithmetic but extended with support for half-precision floating point arithmetic. 

\subsection{Level 1 Workloads}
Level 1 benchmarks include basic parallel algorithms which are common tasks in parallel computing and often used in kernels of real applications. While these applications represent a subset of the Berkeley dwarfs, they are complex enough to represent real applications domains.

\paragraphbe{GUPS} stands for Giga-updates per second. It measures how frequently a computer can issue updates to randomly generated RAM locations. This benchmarks stresses the latency and bandwidth of the device. This test is important because the random memory performance directly maps to the application performance. This is adapted from the HPCC\cite{Luszczek:2006:HCB:1188455.1188677} benchmark suite, 
extended to simplify the tuning of DRAM footprint.

\paragraphbe{Breadth First Search} 
measures the performance for breadth-first search, a
common graph traversal algorithm. This application was
included because it is control-flow intensive. This benchmark is 
adapted from Rodinia, extended with modern CUDA feature support.

\paragraphbe{General Matrix Multiply} is an application that
measures the performance for different types of matrix
multiplications. The types of matrix multiplications include
single and double precision tests with and without transposing
the input matrices. This benchmark is adapted from SHOC with
added support for half precision arithmetic, Tensor Cores, and new CUDA features.

\paragraphbe{Pathfinder} performs a shortest-path algorithm which serves as a test of irregular
parallelism. While most conventional parallel algorithms
have uniform behaviors across the different threads, irregular
algorithms are characterized by different threads performing
different executions. Depending on graph connectivity, different threads can experience
unique behaviors. In addition to this, pathfinder
will experience much higher control flow unit utilization
compared to regular parallelism algorithms as each thread
needs to decide how to execute independently. This benchmark is adapted from Rodinia, extended with new CUDA feature support.

\paragraphbe{Sort} performs fast radix sort \cite{5161005} on an array
of integers. This benchmark was originally included in SHOC. In \ALTIS{}, the workload is extended to simplify dataset size tuning and new CUDA feature support.

\subsection{Level 2 Workloads}
Level 2 benchmarks are macro-benchmarks: real-world application kernels.
Benchmarks in this level are applications that can be
found in industry, and therefore represent a variety of GPU
application domains. These applications represent a diverse types of performance characteristics.

\paragraphbe{CFD Solver} is a computational fluid dynamics benchmark.
This application solves the three-dimensional Euler equations
for compressible flow. This workload optimizes effective
GPU memory bandwidth by reducing total global memory
accesses and overlapping computation. This is adapted from Rodinia, extended to add new CUDA feature support.

\paragraphbe{GPUDWT} is for discrete wavelet transform,
an image and video compression algorithm that is also a
popularly used digital signal processing technique. This
benchmark implements both forward and reverse, as well as
9/7 and 5/3 transforms. The 9/7 transform uses floats while
the 5/3 transform uses integers, so it’s important to measure
the performance for both. This benchmark is adapted from Rodinia, extended to add new CUDA feature support.

\paragraphbe{KMeans} is a popular clustering algorithm used in
data mining. This algorithm shows a
high degree of data parallelism. At the beginning, K centers
are chosen randomly. In each iteration, each data point is assigned
to a center, and at the end of each iteration, each center is
recomputed as the mean of all the data points in its cluster
until the two converge. This benchmark provides 11 different implementations,
including both CPU and GPU side aggregation.
It is extended to add new CUDA feature support.

\paragraphbe{LavaMD} calculates N-body
particle interaction. The code calculates particle potential
and relocation due to mutual forces between particles within
a large 3D space. This space is divided into cubes, or large
boxes, that are allocated to individual cluster nodes. The
large box at each node is further divided into cubes, called
boxes. 26 neighbor boxes surround each box (the home
box). Home boxes at the boundaries of the particle space
have fewer neighbors. Particles only interact with those other
particles that are within a cutoff radius since ones at larger
distances exert negligible forces. Thus the box size is chosen
so that the cutoff radius does not span beyond any neighbor box
for any particle in a home box, thus limiting the reference
space to a finite number of boxes. This benchmark is implemented from scratch and provides 11 different variants, extended to add new CUDA feature support.

\paragraphbe{Mandelbrot} computes an
image of a Mandelbrot fractal, a self repeating geometric
pattern that loops back on itself at ever decreasing sizes.
A commonly used algorithm
is the Escape Time Algorithm, which calculates the value
for different pixels on a per pixel basis. This benchmark was added specifically to test With Dynamic
Parallelism, the benchmark
switches to using the Mariani-Silver Algorithm. Unlike
Escape Time, this procedure starts out coarse grained, and
only iterates at a finer resolution if necessary for certain
subsections. The implementation is adapted from~\cite{mandelbrot}.

\paragraphbe{Needleman-Wunsch} is a nonlinear global optimization
method for DNA sequence alignments. The potential pairs
of sequences are organized in a 2D matrix. In the first step,
the algorithm fills the matrix from top left to bottom right,
step-by-step. The optimum alignment is the pathway through
the array with maximum score, where the score is the value
of the maximum weighted path ending at that cell. Thus,
the value of each data element depends on the values of its
northwest-, north- and west-adjacent elements. In the second
step, the maximum path is traced backward to deduce the
optimal alignment. The benchmark is adapted from Rodinia, extended to add new CUDA feature support.

\paragraphbe{ParticleFilter} is a statistical estimator of the location
of a target object given noisy measurements of that target’s
location and an idea of the object’s path in a Bayesian
framework. The PF has a plethora of applications ranging
from video surveillance in the form of tracking vehicles,
cells and faces to video compression. This particular
implementation is optimized for tracking cells, particularly
leukocytes and myocardial cells. The benchmark is adapted from Rodinia, extended to add new CUDA feature support.

\paragraphbe{SRAD} is a computer vision
application used for reducing noise, or “speckles”, in images
without destroying important image features. This is done
using partial differential equations. Since each stage of this application operates on the
entire image, SRAD requires synchronization after each
stage. This makes SRAD the ideal benchmark to test the
performance of using cooperative groups in CUDA. This benchmark is adopted from Rodinia with \emph{added support for Cooperative Groups}.

\paragraphbe{Where} is a new relational algebra benchmark developed for \ALTIS{}. 
GPUs are increasingly popular for data analytics because relational algebra operations are amenable 
to efficient GPU parallelization~\cite{dandelion}. This benchmark implements a filter for a set of records, returning a subset of the
input records that meet a set of conditions. 
It first maps each entry to a 1 or 0,
before running a prefix sum and using both of these auxiliary
data structures to reduce the input data to just the matching
entries.

\paragraphbe{Raytracing.}
Ray tracing is a rendering technique used for generating images by tracing the path of light in the form of pixels. It operates by simulating the effects of its interaction with virtual objects. This workload is new in \ALTIS{} and the implementation is adapted from 
publicly available implementation here~\cite{raytracing-source}. In addition, an OptiX version~\cite{optix-ray} is supplied for RT Core benchmarking.


\begin{figure*}
 \centering

 \begin{minipage}{.05\textwidth}
 \vspace{-1cm}
  \includegraphics[width=\textwidth]{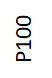} 
  \end{minipage}
 \begin{minipage}{.92\textwidth}
 \centering
  \includegraphics[width=\textwidth]{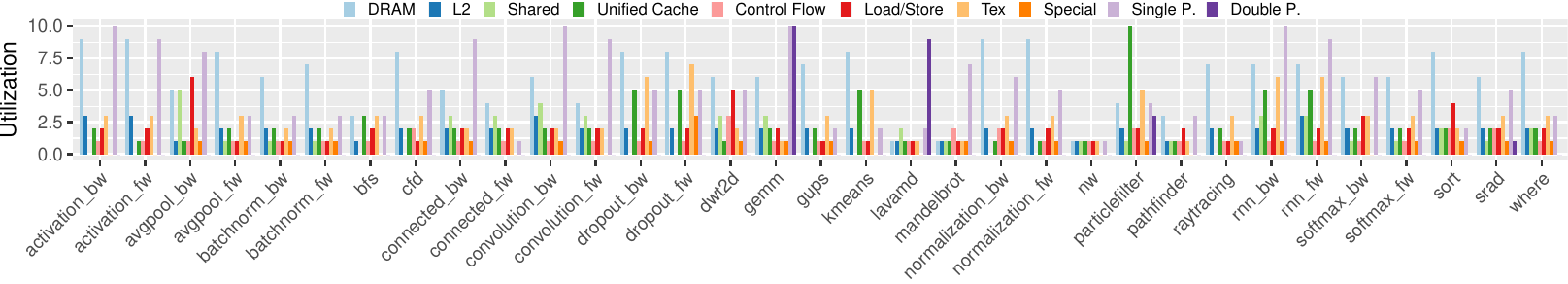}
\end{minipage}
 \begin{minipage}{.05\textwidth}
  \includegraphics[width=\textwidth]{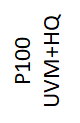} 
 \vspace{-1cm}
  \end{minipage}
 \begin{minipage}{.92\textwidth}
 \centering
  \includegraphics[width=\textwidth]{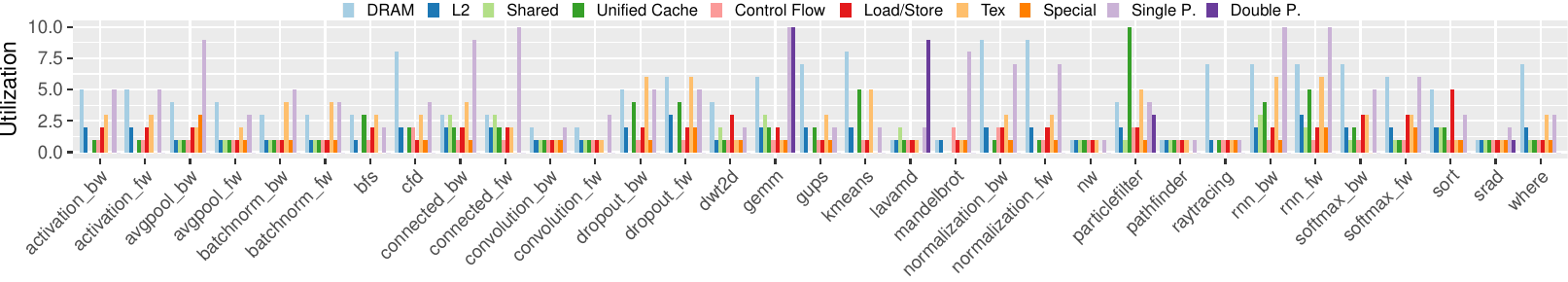} 
  \end{minipage}
 \begin{minipage}{.05\textwidth}
 \vspace{-1cm}
  \includegraphics[width=\textwidth]{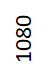}  
 \end{minipage}
 \begin{minipage}{.92\textwidth}
 \centering
  \includegraphics[width=\textwidth]{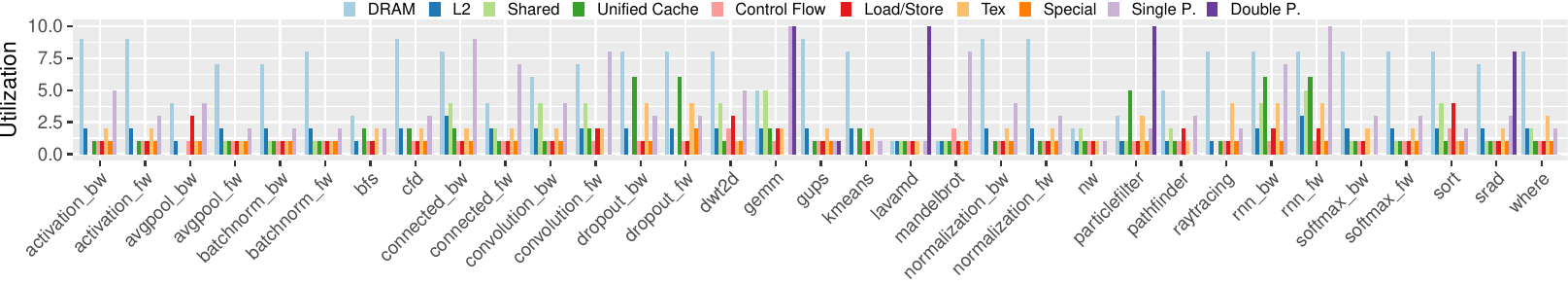}
  \end{minipage}
 \begin{minipage}{.05\textwidth}
 \vspace{-1cm}
  \includegraphics[width=\textwidth]{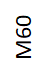}  
 \end{minipage}
 \begin{minipage}{.92\textwidth}
 \centering
  \includegraphics[width=\textwidth]{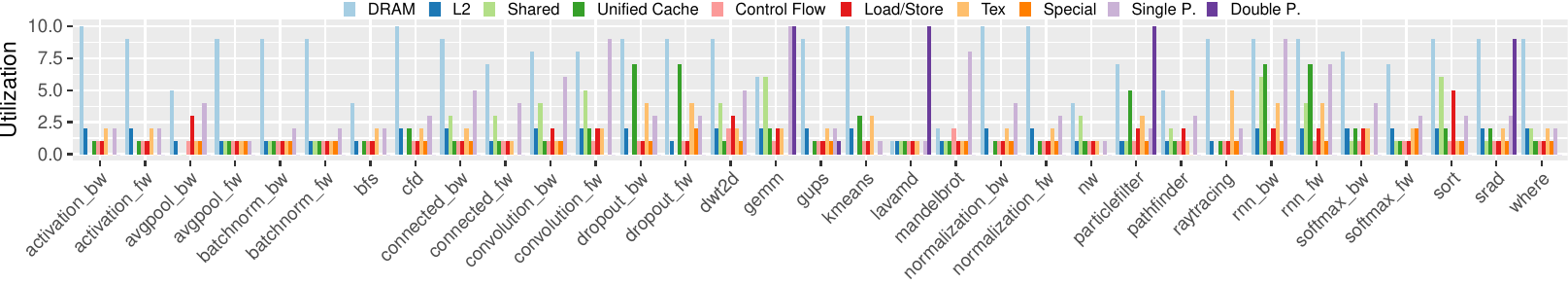}
  \end{minipage}
  \caption{Per-resource type Utilization of \ALTIS{} workloads using P100, GTX 1080, and M60 GPUs.}
  \label{fig:fwd}
  \label{fig:fwd-uvm}
  \label{fig:bwd}
  \label{fig:mirovia}
\end{figure*}

\subsection{DNN Kernel Workloads}
All benchmarks in this section represent  neural network layers commonly seen in popular DNN models. All layers in this section include both forward and backward passes.

\paragraphbe{Activation} layer is used to decide whether a neuron should be activated by calculating the weighted sum and adding bias with it. It introduces non-linearity into the output of a neuron. Some of the most commonly used activation functions include ReLU, sigmoid, tanh, and LeakyReLU. Here we present ReLU since it is the simplest one to understand. ReLU activation function can represented as $y_1=max\{0,x_1\}$ where $x_i$ represents the input to the neuron and $y_i$ is the output.

\paragraphbe{Pooling} is common used between successive convolution layers in a ConvNet architecture. Its main function is to reduce the spatial dimensions on a convolution neural network. For example, applying a maxpool kernel of size $2\times 2$ on a $2\times2$ matrix will yield the biggest number in the matrix, while an average pool kernel of the same size will produce the average value of the sum in the matrix. For simplicity, We include only average pool layer in the report.

\paragraphbe{Batch normalization} is a technique proposed to solve covariate shift \cite{Ioffe2015BatchNA} in DNNs. When parameters in the preceding layer change, the input to the current layer will change accordingly, causing the current layer to adjust to the new distribution. The main goal of batch normalization is to limit the shifting to a certain range to speedup training process and produce reliable models.

\paragraphbe{Connected} layers are those whose neurons are connected to every neuron in the next layer. The connected layer can be seen as a feature vector that holds aggregated information from the previous layer. For example, a connected layer can be right after a convolution layer which provides a low-dimensional in-variate feature space. The fully connected layer can then learn a function from that space to produce more useful or abstract knowledge.

\paragraphbe{Convolution} layer is mostly used to extract important features from images by assigning learning weights to various objects in those images. For example, give an RGB image of size $228\time228$ with 3 channels, we can train a convolution kernel of size $3\times3$ with $3$ channels and stride $1$ to produce an output tensor of size $226\times 226$ with $1$ channel. The output tensor represents one feature in the image, such as the presence of curves in difference parts of the input image.

\paragraphbe{Dropout} is a regularization technique used to prevent neural networks from over-fitting\cite{JMLR:v15:srivastava14a}. The key idea is to randomly drop units from the neural network during training. When training large neural networks on small data sets, over-fitting can be a huge issue when the model is evaluated on test data set. Dropout solves this problem by stochastically introducing noise to prevent units from co-adapting too much, thus making the model more robust.

\paragraphbe{RNN} stands for Recurrent Neural Network. It is widely adopted in learning tasks dealing with sequential data, such as speed recognition, text generation, and so on. RNNs have proven to be successful in capturing the dynamics of sequences by keeping internal states(memory) which tracks information from previous time stamps. Among the most commonly used RNNs are GRU and LSTM. In our benchmark, we only show results for LSTM for simplicity. 

\paragraphbe{Softmax} layer is typically seen as the final output layer in a neural network to perform multi-class classification. It takes an input, usually a score value($z_i$, $i=1...K$), and recomputes it as probabilities. Therefore, the outputs of the layer will represent a true probability distribution, where the sum of each individual output will equal to 1. Its calculation process is shown below:
\begin{equation}
    \sigma{(z_c)}= \frac{e^{z_c}}{\sum_{k=1}^{K}e^{z_k}}
\end{equation}

\paragraphbe{LRN} (Local Response Normalization) is intended to simulate a form of lateral inhibition\cite{NIPS2012_4824} inspired by the type found in real neurons. It allows diminishing response values uniformly large to neighborhoods and creates a high contrast in activation map. This feature is especially useful in unbound activation functions such as ReLU. The original formula is written as
\begin{equation}
    b^{i}_{x,y}=a^{i}_{x,y}/(k+\alpha \sum^{j=min(N-1,i+n/2)}_{j=max(0,i-n/2)}(a^{i}_{x,y})^{2})^{\beta}
\end{equation}
where $b^{i}_{x,y}$ is the regularized output for kernel $i$ at position $x,y$, $a^{i}_{x,y}$ is the source output of kernel $i$ applied at position $x,y$, $N$ is the number of kernels, $n$ is the size of the normalization neighbourhood, and $\alpha, \beta, k$ are hyper parameters of LRN.

\section{Evaluation}
\label{sec:eval}

In this section, we evaluate the applications in \ALTIS{} in terms of runtime characteristics, diversity, and performance. 

\begin{figure}
 \centering
  \includegraphics[width=\columnwidth]{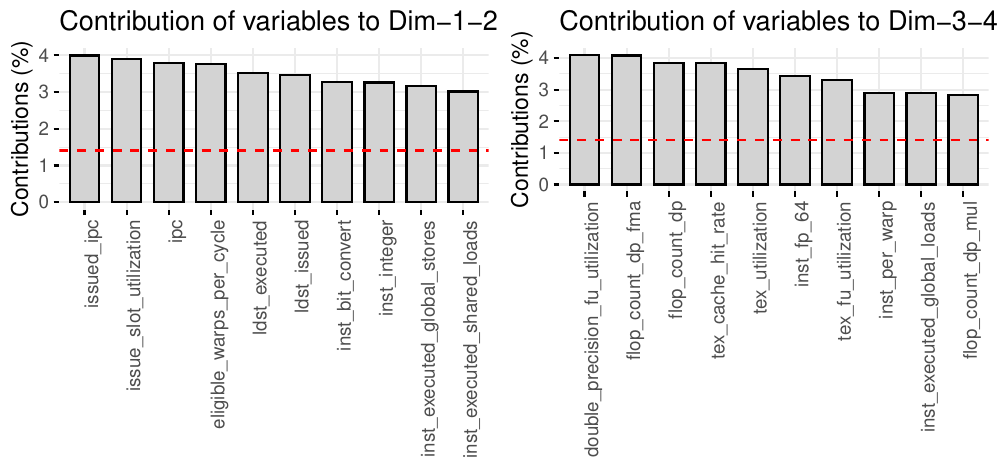}
  \caption{Contribution of the top 10 (out 69) Variables to PCA dimensions 1,2,3, and 4.}
  \label{fig:dim_contrib}
\end{figure}

\subsection{Experimental Setup}
All  measurements are obtained by executing the applications on real hardware. The benchmarks are evaluated on a NVIDIA Tesla P100 GPU with 1.48 GHz shader clock. The CPU is a Intel Xeon E5-2650 running at 2.2 GHz. We use this setup as the standard platform to collect all metrics. We used NVIDIA driver version 418.87 and CUDA version 10.0. The benchmark is also executed on a NVIDIA Tesla M60 running at 1.18 GHz and a NVIDIA GeForce GTX 1080 running at 1.85 GHz to collect utilization metrics. The operating system is Ubuntu 18.04.3 LTS.


\begin{figure}
 \centering
  \includegraphics[width=0.85\columnwidth]{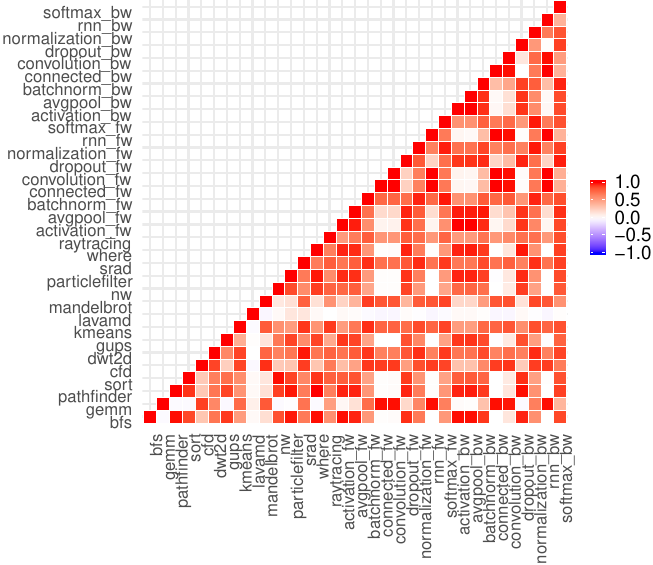}
  \caption{Pearson correlation matrix for Altis}
  \label{fig:corr2}
\end{figure}

\subsection{Benchmark Performance}
We use \ttt{nvprof} to collect the metrics of individual kernels. Note a number of benchmarks involve multiple kernels and some may be redundant. In such cases, we select the maximum utilization of each kernel. The memory and computational unit utilization rate are provided in \autoref{fig:fwd}. 
The utilization rates of different GPU components show a diverse set of behaviors for both forward and backward passes in DNN section of the benchmark. We observe that the most utilized components are DRAM and single precision floating point function unit.

\begin{figure}
    \centering
    \includegraphics[width=0.8\columnwidth]{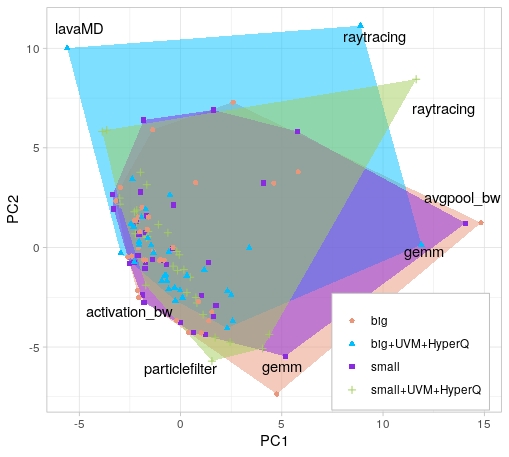}
    \caption{Altis PCA. Blue and gray represent \ALTIS{} running on small and large input datasets respectively.
}
    \label{fig:altis_pca}
\end{figure}

\autoref{fig:altis_pca} and \autoref{fig:dim_contrib} show the \ALTIS{} workloads in  PCA space, as well as the contribution of the top 10 metrics to the first through fourth dimensions of that space. 
We selected counters by performing PCA analysis over all the counters supported by \texttt{nvprof}, and selecting with preference for the most dominant metrics. Because \texttt{nvprof} and GPGPU-sim provide different metrics, we preferred the metrics in \texttt{nvprof} for which there was a clear mapping to a corresponding metric in GPGPU-sim. The major GPGPU kernel characteristics can be categorized into several types: efficiency and utilization, arithmetic features, stall behavior, instruction mix, and memory hierarchy behaviors.  The complete set of metrics used is shown in \autoref{tab:metrics}.

The PCA data show that use of new CUDA features and larger inputs can significantly affect the position of a benchmark in the space, reflect the fact that bottleneck components change when emerging features are exercised.
For example, \texttt{lavaMD} is an outlier in all cases because it uses double-precision units rarely exercised in other workloads, but use of UVM shifts the bottleneck to pipeline stalls. The \texttt{raytracing} and \texttt{nw} workloads behave similarly.

The IPC-related metrics contribute the most to the variance in PC1 while double precision functional units is more prevalent in PC2. This is because only a few benchmarks use double precision data type as indicated in \autoref{fig:fwd}. The new workloads and extended versions of previous workloads enable \ALTIS{} to better cover the PCA space. \ALTIS{}'s new workloads, \ttt{raytracing}, 
and many of the DNN kernels are clearly at extrema of the space.

\begin{table}[htb]
  \centering
    \resizebox{\columnwidth}{!}{
  \begin{tabular}{|c|l|} 
 \hline
 \thead{Category} & \thead{Metrics} \\
 \hline\hline
  Util \& Efficiency & branch\_efficiency,  warp\_execution\_efficiency, warp\_nonpred\_execution\_efficiency, \\
 & inst\_replay\_overhead, gld\_efficiency, gst\_efficiency, ipc, issued\_ipc, issue\_slot\_utilization,\\ 
 & sm\_efficiency, achieved\_occupancy, eligible\_warps\_per\_cycle,ldst\_fu\_utilization,\\
 & cf\_fu\_utilization, tex\_fu\_utilization, special\_fu\_utilization\\
 \hline
 Arithmetic & inst\_integer, inst\_fp\_32, inst\_fp\_64,inst\_bit\_convert,flop\_count\_dp, flop\_count\_dp\_add,\\
 & flop\_count\_dp\_fma,flop\_count\_dp\_mul,flop\_count\_dp\_mul, flop\_count\_sp,flop\_count\_sp\_add,\\
 & flop\_sp\_efficiency,flop\_count\_sp\_fma, flop\_count\_sp\_mul, flop\_count\_sp\_special,\\
 & single\_precision\_fu\_utilization, double\_precision\_fu\_utilization\\
 \hline
 Stall& stall\_inst\_fetch, stall\_exec\_dependency, stall\_memory\_dependency, stall\_texture,stall\_sync,\\
 & stall\_constant\_memory\_dependency, stall\_pipe\_busy,stall\_memory\_throttle, stall\_not\_selected\\
 \hline
 Instructions & inst\_executed\_global\_loads, inst\_executed\_local\_loads, inst\_executed\_shared\_loads, \\ 
 & inst\_executed\_local\_stores, inst\_executed\_shared\_stores, inst\_executed\_global\_reductions, \\
 & inst\_executed\_tex\_ops, l2\_global\_reduction\_bytes, inst\_executed\_global\_stores, \\
 & inst\_per\_warp, inst\_control, inst\_compute\_ld\_st, inst\_inter\_thread\_communication,\\
 & ldst\_issued,ldst\_executed.\\
 \hline
 Cache\&Mem & local\_load\_transactions\_per\_request,global\_hit\_rate, local\_hit\_rate, tex\_cache\_hit\_rate, \\
 & l2\_tex\_read\_hit\_rate, l2\_tex\_write\_hit\_rate, ram\_utilization, shared\_efficiency,\\
 & shared\_utilization, l2\_utilization, tex\_utilization, l2\_tex\_hit\_rate.\\
 \hline
\end{tabular}
}
\caption{Metrics used to create Altis' PCA metric space.
}
\label{tab:metrics}
\end{table}

\begin{figure}
    \centering
    \includegraphics[width=.8\columnwidth]{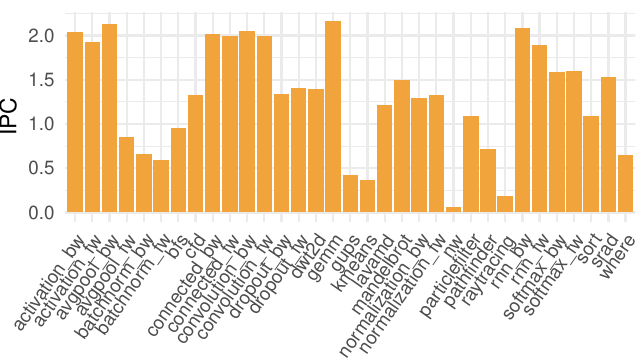}
    \caption{IPC for \ALTIS{} measured with the largest supported data set size.}
    \label{fig:ipc}
\end{figure}

The floating point function unit is closely related to the IPC for most kernels. For example, convolution is compute intensive, which results in high IPC, shown in \autoref{fig:ipc}. Low utilization of single precision function unit results in low IPC for batch normalization. The eligible number of warps per cycle also shows high number for convolution and low value for batch normalization. This can be explained by that convolution has relatively good data locality, whereas batch normalization requires more memory operations which reduces the number of warps eligible to issue the next instruction. This is a sign that convolution operation is compute bound and batch normalization is memory bound.

\begin{figure}
    \centering
    \includegraphics[width=.8\columnwidth]{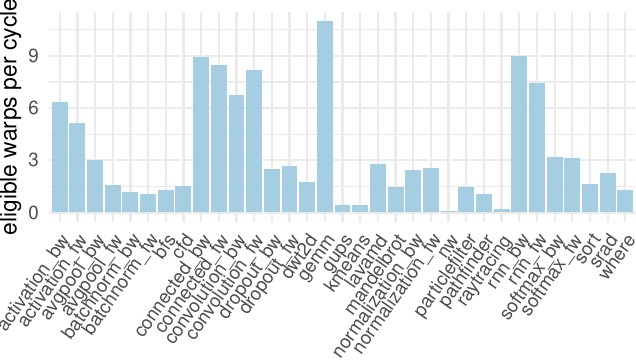}
    \caption{\ALTIS{} Average Eligible Warps Per Cycle}
    \label{fig:eligible_warps_per_cycle}
\end{figure}

\autoref{fig:eligible_warps_per_cycle} shows the eligible warps per cycle in \ALTIS{}.    
Eligible warps tend to show how often the benchmark makes data requests, and it correlates with IPC. For example, \texttt{gemm} and \texttt{connected\_fw} are heavily computation bound since they are essentially matrix-matrix multiplication. In contrast, \texttt{gups} always requests a single (randomly chosen) unit of data from DRAM for each read, and the resulting stalls result in very low eligible warps per cycle.
    
The utilization for the rest of all benchmarks in \ALTIS{} show a diverse range of values. Each GPU component utilization is increased compared to SHOC. This can be explained by the increase in input data size, 
which demonstrates the importance of having user-defined input problem size to stress hardware performance. These benchmarks also differ from DNN kernels. DNN kernel tend to stress dram and single precision function units heavily, while the conventional benchmarks exhibit a more diverse utilization of each component. Compared to \autoref{fig:corr1}, \autoref{fig:corr2} shows a great deal of variance in the correlation between benchmarks. The strong correlation between \texttt{gemm} and convolution kernels indicates they share similar characteristics as both are compute bound. On the other hand, gups has almost no correlation with convolution, since random memory access is heavily memory bound. The correlation matrix shows a good amount of applications with little correlation, indicating diverse GPU behaviors. In addition, relative to the utilization reported in \S\ref{sec:related} for Rodinia and SHOC, the hardware is much more fully utilized, and the majority of workloads have at least one resource whose utilization is a significant fraction of peak. 

\paragraphbe{Discussion: Insights from \ALTIS{}.} Overall, the data illustrate \ALTIS{}'s primary value in providing new workloads sized and crafted to \emph{actually utilize} the hardware and \emph{exercise new features}. While DRAM and floating point are known to be critical to GPU performance, and \ALTIS{} will not change that, exercising new features at meaningful scale has revealed new insights. For example, the first order bottlenecks for some workloads have changed, as \autoref{fig:altis_pca} and \autoref{fig:fwd} illustrate. UVM may decrease performance for some workloads, but increases utilization under several metrics: while this observation has been made by others, a shared suite that exercises the feature in an understandable way prevents researchers from having to directly modify suites in potentially different ways to model UVM, as was necessary for works such as \cite{mosaic17micro, tianhao-hpca16}. DRAM utilization, single precision functional unit utilization, and unified cache utilization are all metrics that show significant changes in behavior in \ALTIS{} compared to the original  versions. As we show in per-feature analysis below, new features such as Cooperative Groups have unpredictable and undesirable performance artifacts: supporting them in \ALTIS{} enables research into how to improve them. 

\subsection{CUDA Feature Analysis}
In this section, we analyze a subset of benchmarks with new CUDA features to explore each feature's impact.

\paragraphbe{Unified Memory}: We measure kernel time
plus transfer time of BFS without UVM and compare to kernel time with UVM,
since there is no explicit transfer time with UVM. Three different versions of BFS using UVM were measured and  compared to the baseline with no new CUDA features.
The first version uses UVM without
\texttt{cudaMemAdvise()} or \texttt{cudaMemPrefetchAsync()}.
The second uses only
\texttt{cudaMemAdvise()}, and the last uses both \texttt{cudaMemAdvise()} and \texttt{cudaMemPrefetchAsync()}.
\begin{figure}[h!]
 \centering
  \includegraphics[width=.75\columnwidth]{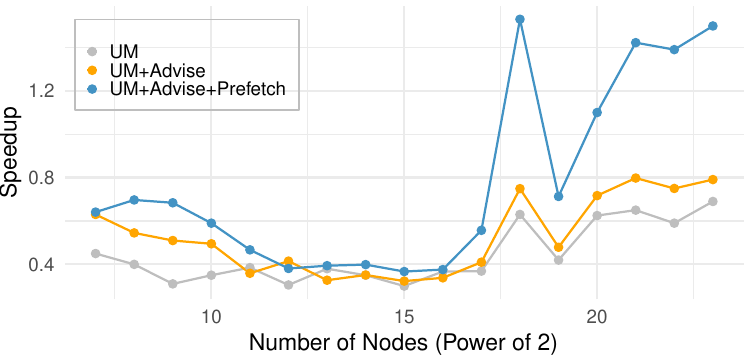}
  \caption{BFS Speedup Using Unified Memory}
  \label{fig:unified_mem}
  \vspace{-.2cm}
\end{figure}
BFS with UVM is faster than the baseline version only with pre-fetching enabled. Additionally, the speedup was inconsistent and did not scale with the input size. This is because the
execution path is highly dependent on the generated graph. While data is randomly generated, irregular access patterns typified by graph workloads are a challenge for UVM because hiding frequent demand paging latency depends on prefetch that is not always effective.

\begin{figure}[h!]
 \centering
  \includegraphics[width=.75\columnwidth]{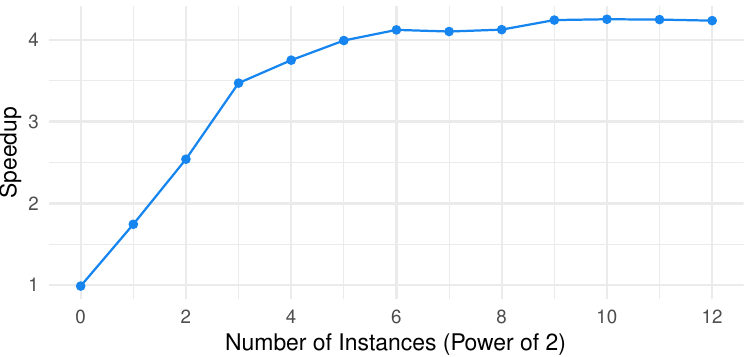}
  \caption{Pathfinder Speedup Using HyperQ}
  \label{fig:hyperq}
  \vspace{-.2cm}
\end{figure}

\paragraphbe{HyperQ}: We explore HyperQ impact on Pathfinder. HyperQ increases SM utilization
when multiple independent kernels can execute concurrently.
Our updated pathfinder runs multiple instances on different
streams. \autoref{fig:hyperq} shows the speedup as the
number of concurrent kernels increases. Transfer time is elided because it is constant.

Speedup from HyperQ increases with the number of parallel kernels, and
levels out around 32 instances, when it saturates all 32 
work queues supported by the physical hardware. 
We see speedup starting
at a little under 1x for a single instance, and up
to 4x thereafter. This occurs because increasing the number of
instances increases queue occupancy, but aggregate throughput becomes limited by available SMs. 

\paragraphbe{Cooperative Groups}: We measure and compare
SRAD using cooperative kernels 
to the kernel time for the original implementation.
Transfer time is elided.
\begin{figure}[h!]
 \centering
  \includegraphics[width=.75\columnwidth]{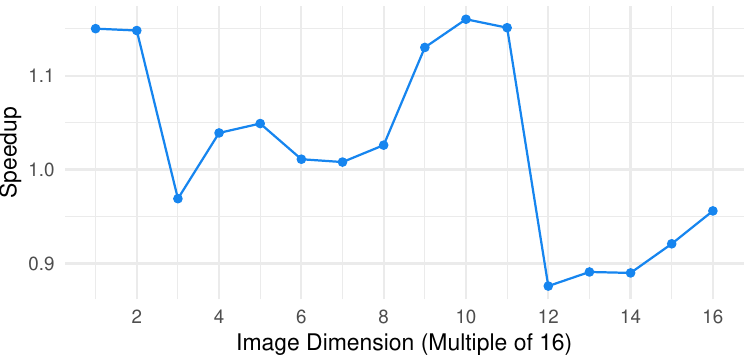}
  \caption{SRAD Speedup Using Cooperative Groups}
  \label{fig:coop}
  \vspace{-.2cm}
\end{figure}
A key concern for
cooperative groups is the
limit on the number of blocks the GPU is able to launch. SRAD using a cooperative kernel could not be run on
image sizes greater than 256x256. We vary the problem size by multiples of 16. The data in \autoref{fig:coop} show that the feature 
provides minimal performance benefit in a handful of cases, and can harm 
performance significantly in others. Supporting this feature in a way
that enables developers to reason about and predict the utility of the feature is possible direction for research exposed by \ALTIS{}.

\paragraphbe{Dynamic Parallelism}: For this feature, the speedup
was measured using the kernel times for Mandelbrot with
and without Dynamic Parallelism. Transfer time was not included.
The benchmark shows smooth increase in
speedup as problem sizes increase. This is due to the efficiency of the algorithms used and the more efficient coding of the problem that
Dynamic Parallelism enables. While the traditional Escape
Time algorithm is forced to calculate values for every pixel,
Mariani-Silver can subdivide and thus ignore ever
increasing swaths of the image without requiring synchronization
or relaunch of multiple kernels. 
\begin{figure}[h!]
 \centering
  \includegraphics[width=.75\columnwidth]{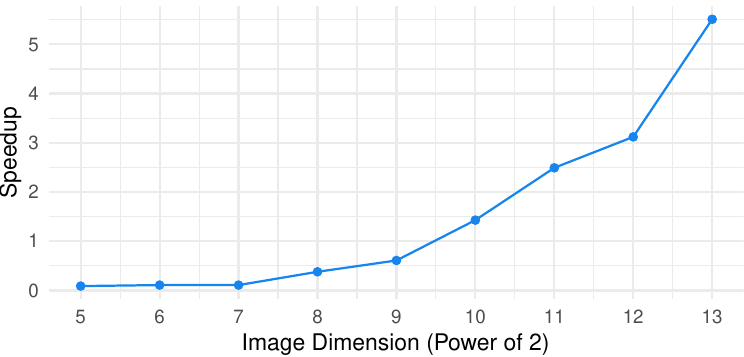}
  \caption{Mandlebrot Speedup using Dynamic Parallelism}
  \label{fig:dynamic}
  \vspace{-.2cm}
\end{figure}

\begin{figure}[h!]
 \centering
  \includegraphics[width=.75\columnwidth]{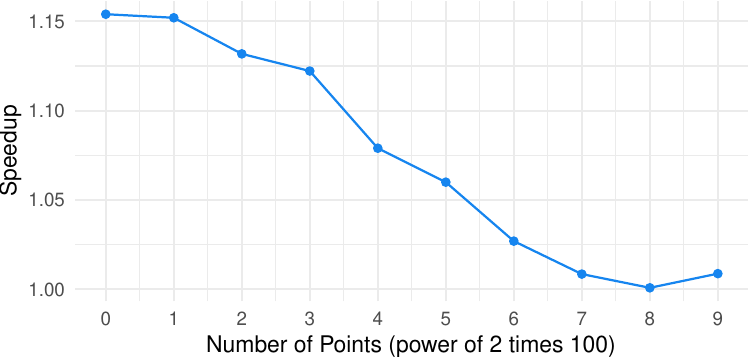}
  \caption{Particlefilter Speedup Using CUDA Graph}
  \label{fig:cu_graph}
  \vspace{-.2cm}
\end{figure}

\paragraphbe{CUDA Graph}: For this feature, the speedup
was measured using the frame processing times for particlefilter
with and without CUDA Graph.
The frame dimension is set to $30 \times 30$, the frame is set to $40$.
Data transfer time and data initialization time are not included. The data in \autoref{fig:cu_graph} demonstrates slight speedup
as number of points increases.
By defining work as graphs, all the kernel work descriptors
are pre-initialized and can be launched repeatedly as rapidly as possible, thus reducing
launching overheads.
As the data size increases, the kernel launch time
is overshadowed by the computation time, thus less speedup.

\section{Conclusion}
Our goal with \ALTIS{} is to modernize GPGPU benchmark suites. We improved the diversity over existing benchmarks, introducing
new workloads from different domains, adapting
problem sizes to better match modern hardware, and adding 
support for new CUDA features.  
We hope \ALTIS{} will server as a more complete GPGPU benchmark suite for modern GPGPU research.

\section{Acknowledgements}

We thank the PC reviewers and our shepherd for their valuable feedback.  We thank Christopher Denny, Sarah Wang, and Vance Miller, who contributed to the engineering effort to build many of the workloads in \ALTIS{}. This research was supported by NSF grants CNS-1618563 and CNS-1846169.

\bibliography{rodinia,clean_uberbib} 
\bibliographystyle{ieeetr}
\balance
\end{document}